\documentclass{article}
\usepackage{caption}
\usepackage{subcaption}
\usepackage{authblk}
\usepackage{amsmath}
\usepackage{graphicx}
\usepackage{amssymb}

\usepackage{amsmath}
\usepackage{bm}
\usepackage{siunitx} 

\begin{document}

\title{Neural Operator Modeling of Platelet Geometry and Stress in Shear Flow}

\author[1]{Marco Laudato\footnote{Corresponding authors; laudato@kth.se}}
\author[2]{Luca Manzari}
\author[3]{Khemraj Shukla}
\affil[1]{FLOW Research Center, Department of Engineering Mechanics, KTH Royal Institute of Technology, Stockholm, SE-10044, Sweden.}
\affil[2]{PDC Center for High Performance Computing, KTH Royal Institute of Technology, Stockholm, SE-11428, Sweden}
\affil[3]{Division of Applied Mathematics, Brown University, Providence, RI 02906, USA}

\maketitle

\begin{abstract}
Thrombosis involves processes spanning large-scale fluid flow to sub-cellular events such as platelet activation. Traditional CFD approaches often treat blood as a continuum, which can limit their ability to capture these microscale phenomena. In this paper, we introduce a neural operator-based surrogate model to bridge this gap. Our approach employs DeepONet, trained on high-fidelity particle dynamics simulations performed in LAMMPS under a single shear flow condition. The model predicts both platelet membrane deformation and accumulated stress over time, achieving a mode error of $\sim0.3$\% under larger spatial filtering radii.
At finer scales, the error increases, suggesting a correlation between the DeepONet architecture’s capacity and the spatial resolution it can accurately learn. These findings highlight the importance of refining the trunk network to capture localized discontinuities in stress data. Potential strategies include using deeper trunk nets or alternative architectures optimized for graph-structured meshes, further improving accuracy for high-frequency features. Overall, the results demonstrate the promise of neural operator-based surrogates for multi-scale platelet modeling. By reducing computational overhead while preserving accuracy, our framework can serve as a critical component in future simulations of thrombosis and other micro-macro fluid-structure problems.

\end{abstract}

\section{Introduction}
\label{sec:intro}
Multi-scale modeling is critical in computational fluid dynamics (CFD) and particle dynamics, especially for complex biomedical flows like blood clotting in vessels~\cite{laudato2024high}. Traditional CFD treats blood as a continuum~\cite{laudato2023buckling,laudato2024analysis,laudato2024sound}, but this approach cannot capture cellular-scale phenomena such as platelet activation and aggregation~\cite{zhang2021predictive}. For instance, platelet activation involves filopodial extensions and other molecular-scale events that continuum models cannot describe. Coupled micro–macro simulations represent a relatively new frontier for the description of this kind of phenomena. Pioneering work by Bluestein and colleagues integrated Dissipative Particle Dynamics (DPD) for macroscopic blood flow with coarse-grained molecular dynamics (CGMD) for individual platelets, using hybrid interfaces and multiple time-stepping to bridge disparate scales. Such multi-scale models provide mechanistic insight into thrombosis, but they are computationally intensive and cannot be effectively scaled to a continuous multi-phase description of blood flow. Recent years have therefore seen a resurgence of data-driven and machine learning (ML) methods in fluid mechanics. In cardiovascular simulations in particular, researchers are moving beyond continuum by incorporating ML-driven surrogate models and multi-physics coupling to improve predictive power~\cite{schwarz2023beyond}.

Machine learning offers new avenues to integrate fine-scale physics into coarse-scale simulations more efficiently. One breakthrough is the use of neural operators, i.e. a deep learning model that learns mappings between function spaces, enabling fast surrogates for physical processes. Karniadakis and co-workers introduced DeepONet as a neural operator capable of learning nonlinear solution operators for PDEs~\cite{lu2021learning}. Recently, this approach has been applied to platelet dynamics: Laudato et al. trained a DeepONet on high-fidelity LAMMPS particle simulations of a deformable platelet, allowing it to predict the platelet’s shape response to shear flow~\cite{laudato2024high}.
Remarkably, the learned surrogate reproduces complex membrane deformations with ~0.5\% mode error. Such accuracy is promising to serve as an online replacement for the microscale model, providing a scalable interface between sub-platelet physics and continuum blood flow. Neural operators enable real-time coupling of micro and macro scales that would be prohibitively expensive with direct simulation alone. Similarly, physics-informed neural networks (PINNs) have been employed for multi-scale data assimilation; for example, PINNs were used to infer thrombus material properties non-invasively by embedding clinical observations into the simulation of blood flow and clot mechanics~\cite{cai2021physics}.

Another key contribution of ML is in accelerating multi-scale computations. Bluestein’s group has demonstrated AI-aided multiscale simulation frameworks that significantly reduce computational cost~\cite{zhang2015multiple}. Han et al. (2022) developed an adaptive multi-time-stepping algorithm guided by ML, achieving a 500× speed-up in platelet-based flow simulations~\cite{han2022scalable}. This enabled them to simulate clot formation with over 100 million resolved particles, a previously unattainable spatial–temporal resolution for blood flow models. Such performance gains arise by using learned models to optimize time-step selection and load balancing, illustrating how ML can push multi-scale CFD into the realm of ultra-high resolution. In another study, a biomechanics-informed online learning strategy was used to generalize classic theoretical results (Jeffery’s orbits for rigid ellipsoids) to deformable cells in flow. By learning additional parameters for the cell’s motion from simulation data, the model captures deformability effects on particle trajectories that traditional formulas neglect. This exemplifies how data-driven learning of corrections to continuum theories can bridge micro-scale physics (cell elasticity, shape change) with macro-scale flow behavior.

In turbulent and multiphase flows, ML is increasingly used to derive closure models that embed microscopic physics into macroscopic equations. An interesting example is the work of Siddani and Balachandar, who trained rotation-equivariant neural networks to predict the hydrodynamic forces and torques on particles in dense suspensions~\cite{siddani2023point}. Their ML-based point-particle model achieves up to 85–96\% accuracy in reproducing neighbor-induced force and torque fluctuations, across a wide range of Reynolds numbers and particle concentrations. By learning from fully resolved simulations, these closures effectively transfer microscale interaction physics (wake effects, multi-particle collisions) into a form usable in large-scale CFD. This data-driven approach outperforms traditional low-order closures and extends their validity into regimes that were previously beyond reach.

ML techniques enhance multi-scale modeling by combining underlying mechanistic knowledge while addressing computational bottlenecks. In blood flow simulations, this means that phenomena like platelet activation (once requiring brute-force multi-scale coupling) can now be encapsulated in trained networks that plug into continuum solvers with minimal loss of accuracy~\cite{laudato2024high}. The result is an effective bridge between microscopic and macroscopic scales, enabling simulation of thrombosis and hemodynamics with unprecedented detail and efficiency. This synergy of ML and CFD is rapidly evolving as state-of-the-art approaches now leverage neural surrogates, PINNs, and other algorithms to tackle long-standing multi-scale challenges in fluid mechanics. 

Ultimately, these innovations have the potential to improve predictive modeling of complex fluid–particle systems, from platelet-rich blood flows to turbulent multiphase processes, by capturing fine-scale dynamics without sacrificing tractability.

In this work, we present a DeepONet-based model capable of predicting the dynamic behavior of a deformable platelet in shear flow. Additionally, the model accurately describes the time evolution of accumulated stress on the platelet membrane, a key observable linked to platelet activation~\cite{zhang2021predictive}. The training dataset consists of deformed platelet configurations at different time steps, generated through molecular dynamics simulations in LAMMPS. Our results demonstrate a prediction mode error of approximately 0.3\% compared to ground truth, highlighting the model’s ability to effectively capture microscopic platelet dynamics. Furthermore, we investigate DeepONet’s ability to capture spatial resolution by analyzing its response to discontinuous accumulated stress data from LAMMPS. To this end, we apply two linear filters of varying spatial radii to assess how different filtering windows affect the model’s predictive capability.

In section~\ref{sec:gen_training}, the details of the particle dynamics implementation in LAMMPS to produce the training dataset are discussed. In section~\ref{sec:don}, the DeepONet implementation and training details are presented. In section~\ref{sec:results}, the relative difference between the model and the ground truth for different filter radii are discussed. 

\section{Generating the training dataset}
\label{sec:gen_training}
To generate the learning dataset, we employ Dissipative Particle Dynamics (DPD)~\cite{groot1997dissipative} to capture the time evolution of a simplified platelet under a single shear flow condition. This approach builds on our earlier implementation~\cite{laudato2024high}, where we simulated final platelet shapes across multiple shear stress values but recorded only the end-state configurations. Here, by contrast, we focus on a single shear stress value and track every time step, thus obtaining the complete time evolution of the platelet’s deformation and accumulated stress.

\begin{figure}[h!]
    \centering
    \includegraphics[width=0.67\linewidth]{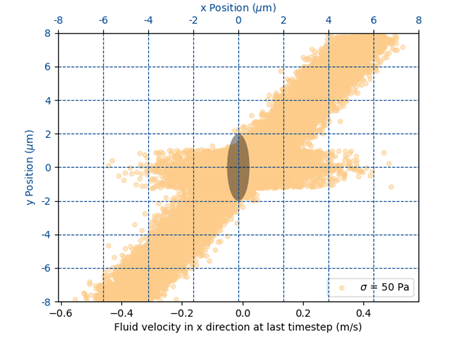}
    \caption{\textit{Diagram of the particle dynamics simulation domain (cross-section at $z=0$). The initial position of the platelet is shown as a grey ellipse. A Couette flow is generated by imposing blood velocity at the boundaries. The velocity dispersion arises from the statistical temperature. The horizontal profile around the platelet is due to the non-penetration boundary conditions which prevent blood particles from entering the platelet’s membrane.}}
    \label{fig:couette}
\end{figure}

The platelet is modeled as a hollow deformable ellipsoid (4×4×1 µm) composed of approximately 18,000 particles, arranged via 3D Delaunay triangulation and connected by harmonic bonds. The spring constant for these bonds is tuned to reproduce experimentally consistent global material properties~\cite{zhang2014multiscale}. The platelet is suspended in a box of dimensions 16×16×8 µm (see~\cite{zhang2014multiscale}), where the lower and upper walls move in opposite directions to generate a Couette flow (see Figure~\ref{fig:couette}). 

For this work, the wall velocities correspond to a shear stress of 50 Pa. Periodic boundary conditions are applied in the remaining directions, while ghost particles near the walls enforce no-slip conditions. The fluid itself is modeled via DPD, with repulsive, dissipative, and random components capturing inter-particle interactions, as well as particle–platelet interactions, thereby allowing the membrane to deform naturally under flow. Additional details of the implementation can be found in~\cite{laudato2024high}.
We use LAMMPS to advance the system with a velocity-Verlet time integrator, and the chosen time step (2.4×$10^{-10}$ s) ensures stability and accuracy of the simulation. Over one Jeffery’s orbit (i.e., the characteristic flipping period of a rigid ellipsoid in shear), we collect the platelet’s configuration at each time step, obtaining a time series of deformed geometries. In addition to the coordinates of the membrane particles, we record an invariant of the instantaneous stress tensor, $\hat{\tau}(p, t)$, where p denotes a given membrane particle and t represents time, defined as~\cite{apel2001assessment}

\begin{equation}
    \hat{\tau}(p,t) = \frac{1}{\sqrt{3}} \sqrt{\tau_{xx}^2+\tau_{yy}^2+\tau_{zz}^2-\tau_{xx}\tau_{yy}-\tau_{xx}\tau_{zz}-\tau_{yy}\tau_{zz}+3(\tau_{xy}^2+\tau_{yz}^2+\tau_{xz}^2)}
\end{equation}

\noindent
Where $\tau_{ij}$ is the stress tensor evaluated for each membrane particle at each timestep during the particle dynamics simulation.

\begin{figure}[h!]
    \centering
    \includegraphics[width=0.67\linewidth]{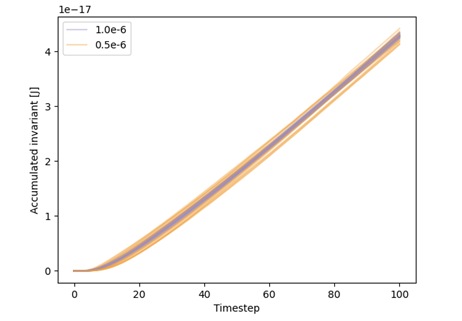}
    \caption{\textit{: The accumulated invariant is plotted versus time for a set of membrane particles. The colors indicate the two filter radii. The DeepONet surrogate model will learn this time evolution for every membrane particle. As expected by the filtering process, the more aggressive filter (purple) results in a smaller dispersion.}}
    \label{fig:invariant}
\end{figure}

\noindent
This quantity is then accumulated over time at each timestep, providing a measure closely linked to platelet activation~\cite{apel2001assessment}. The values of the accumulated stress obtained in this way result discontinuously distributed on the platelet membrane. A common approach in molecular dynamics~\cite{zhang2014multiscale} is to post-process these values with a low-pass spatial filter. To this end, we have employed Paraview’s Point Dataset Interpolator with a linear kernel. In particular, two different radii have been selected (0.5x$10^{-6}$ m, and 1.0x$10^{-6}$ m), to analyze the sensitivity of the neural operator to different spatial resolutions. Figure~\ref{fig:invariant} shows the behavior of the accumulated invariant $\hat{\tau}$ as a function of time for multiple membrane particles. As expected, the spatial dispersion of the values in time is reduced for the more aggressive filter. The resulting datasets serve as the foundation for training and validating the DeepONet-based surrogate model, enabling it to predict both the platelet’s shape evolution and the time-dependent distribution of accumulated membrane stress.

\section{DeepONet implementation}
\label{sec:don}
The neural network architecture used to model the dynamics of the platelet in shear flow is DeepONet, a deep operator network designed to approximate nonlinear continuous operators, as supported by the universal approximation theorem~\cite{lu2021learning}. In this work, DeepONet is trained to approximate the operator $G$, which represents both the displacement and the accumulated stress of each particle in the platelet’s initial configuration $\xi_0$ over time. Given the initial platelet configuration $\xi_0$ and the blood shear $\sigma$, the operator $G(\xi_0, t; \sigma)$ predicts the time evolution of the displacement field and the corresponding accumulated stress distribution for each time step $t$.

\begin{figure}[h!]
    \centering
    \includegraphics[width=0.49\linewidth]{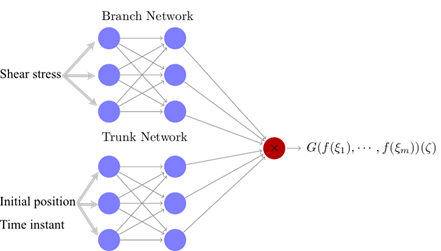}
    \includegraphics[width=0.45\linewidth]{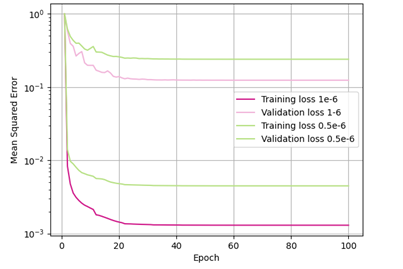}
    \caption{\textit{The left panel shows the general architecture of DeepONet employing FCNs, where $\xi_i$ represents the i-th initial position and time instant and $\zeta$ represents the value of the shear stress. The right panel shows the training and validation losses (mean squared error) for the two different filters. The validation dataset represents 10\% of the total training dataset.}}
    \label{fig:don_training}
\end{figure}

DeepONet consists of two deep neural networks (see Figure~\ref{fig:don_training}, left panel): the branch network, which encodes the boundary and initial conditions as well as the parameters of the PDEs, and the trunk network, which encodes the independent variables where the output functions are evaluated. In this implementation, both the branch and trunk networks are fully connected feed-forward neural networks (FCNs). The branch net takes as input the scalar value of the shear stress $\sigma$, while the trunk net processes the coordinates of the initial platelet configuration and the considered time instant. The branch network consists of 2 inner layers of 32 and 16 neurons, respectively. The trunk network consists of 3 inner layers of 32, 32, and 16 neurons, respectively. The latent space dimension is 32. These meta-parameters have been selected following the sensitivity study in~\cite{laudato2024high}.

The outputs of the branch and trunk networks are then combined through an inner product and passed through a final dense layer. The output layer has four dimensions, representing both the three-dimensional displacement and the accumulated stress of each particle at every time step.
The rectified linear unit (ReLU) activation function is used in both FCNs, while a linear activation function is applied in the output layer to allow the prediction of both positive and negative values. The loss function is the mean squared error (MSE) between the network’s predictions and the corresponding particle dynamics simulation data. As shown in Figure~\ref{fig:don_training} (right panel), training converges within 100 epochs. The neural operator is implemented in Python using the TensorFlow library.

\section{Results and Discussion}
\label{sec:results}
\begin{figure}[h!]
    \centering
    \includegraphics[width=0.67\linewidth]{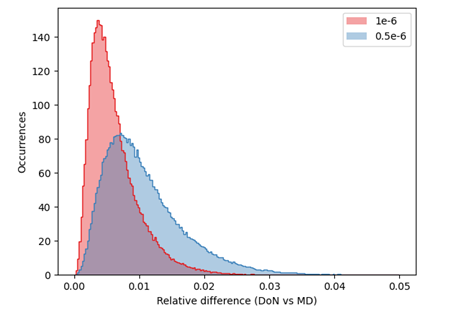}
    \caption{\textit{Relative difference distribution between the DeepONet model prediction and the ground truth evaluated on the validation dataset for the two different spatial filters. The larger value of the filter radius (which produces a more homogeneous spatial distribution of the accumulated stress on the platelet) produces a maximum error of 3\% and a mode error of 0.3\%. The smaller value of the filter produces a maximum error of 4\% and a mode error of 0.7\%.}}
    \label{fig:rel_diff}
\end{figure}

Once trained, the DeepONet-based model can predict both the displacement of the platelet’s membrane and the values of the accumulated stress at a given instant of time. The quality of the prediction, however, depends on the spatial filter’s radius for the values of the accumulated stress introduced in section~\ref{sec:gen_training}. The relative difference $D$ between the network prediction and the LAMMPS simulations is defined as

\begin{equation}
    D=\frac{|| G_{pred}-G_{true}||}{||G_{true}||}
\end{equation}

\noindent
where we have employed the $L^2$ norm. The error distributions of the network predictions when evaluated on a test dataset (not employed during the training) relative to the different filter radii are shown in Figure~\ref{fig:rel_diff}. For the larger value of the filter radius (1.0x10$^{-6}$ m), the error mode is approximately 0.3\% while the maximum error is below $\sim3$\%. For the smaller filter radius (0.5x10$^{-6}$ m), the error mode is about 0.7\% with a maximum error smaller than $\sim4$\%.

Interestingly, these results suggest a possible connection between the neural operator’s capacity (i.e., the number of trainable parameters, depth, and architecture of both the branch and trunk networks) and the finest spatial resolution it can reliably learn. In other words, while DeepONet accurately captures the accumulated stress distributions for larger filter radii, the significant increase in prediction error at the smallest radius hints that the network’s representation power might be insufficient to resolve finer-scale discontinuities. Indeed, if the trunk network in particular cannot adequately represent high-frequency variations in the membrane’s stress or displacement signals, it will struggle to learn or generalize those finer details from the training set. This aspect is especially important for platelet activation models including filopodia extension. Extending the trunk net with additional layers, increasing the number of neurons, or using more sophisticated feature-extraction strategies could alleviate these limitations and improve the model’s accuracy at smaller filter radii.

A promising direction for future research is to modify the trunk net architecture to better handle high-resolution features. For instance, one could employ a Graph Neural Network (GNN) to more naturally encode the platelet’s mesh structure and handle intricate local relationships between membrane particles. Likewise, Fourier Neural Operators (FNOs) which leverage Fast Fourier Transforms to capture both global and local behaviors—could be tested on the same training and test sets to assess whether they can resolve fine-scale stress discontinuities more effectively. While GNNs excel at processing graph-structured data (like a discretized membrane), FNOs have shown promise in modeling complex PDE-driven phenomena in continuous domains. Each approach comes with trade-offs in terms of implementation complexity, computational overhead, and required hyperparameter tuning, but both represent potentially powerful alternatives for improving spatial resolution in DeepONet-based platelet models. Overall, these findings underscore the promise of neural operators as efficient and accurate surrogates for microscale platelet dynamics. Future works will explore multiple shear values, integrate more advanced trunk net designs, and embed this neural operator framework into continuum CFD solvers for a fully multi-scale representation of thrombosis.

\section*{Acknowledgments}
M.L. was supported by Swedish Research Council Grant No. 2022–03032. The numerical computations were enabled by resources provided by the National Academic Infrastructure for Supercomputing in Sweden (NAISS) at the PDC Center for High Performance Computing, KTH Royal Institute of Technology, Sweden, partially funded by the Swedish Research Council through grant agreement no. 2022-06725.


\end{document}